\documentclass[twocolumn,twoside,slac_two]{revtex4}

\usepackage{graphicx}
\usepackage{fancyhdr}
\begin{document}

\title{\boldmath $|V_{ub}|$ from the Spectrum of $B\to\pi e\nu$}

%

\author{Patricia Ball}
\affiliation{IPPP, Department of Physics,
University of Durham, Durham DH1 3LE, UK}

\begin{abstract}
I discuss the results for $|V_{ub}|f_+(0)$ 
and $|V_{ub}|$ obtained from the spectrum of 
$B\to\pi e \nu$ and the form factor $f_+(q^2)$ 
 from QCD sum rules on the light-cone and unquenched lattice
calculations; the shape of $f_+(q^2)$
is fixed from experimental data.
\end{abstract}

\maketitle

\thispagestyle{fancy}

The determination of $|V_{ub}|$ from $B\to\pi\ell\nu$ requires a
theoretical calculation of the hadronic matrix element
\begin{eqnarray}
\lefteqn{\langle \pi(p_\pi)| \bar u \gamma_\mu b | B(p_\pi+q)\rangle
 =}\nonumber\\
&=& \left(
2p_\pi{}_\mu  + q_\mu - q_\mu\,\frac{m_B^2-m_\pi^2}{q^2}\right)
 f_+(q^2)\nonumber\\
&&{} + 
 \frac{m_B^2-m_\pi^2}{q^2}\, q_\mu \,f_0(q^2)\,,
\end{eqnarray}
where $q_\mu$ is the momentum of the lepton pair, 
with $0\approx m_{\ell}^2\leq q^2\leq
(m_B-m_\pi)^2=26.4\,$GeV$^2$. $f_+$ is the dominant form factor,
whereas $f_0$ enters only at order $m_{\ell}^2$ and can be neglected
for $\ell=e,\mu$. The spectrum of $B\to\pi\ell\nu$ in $q^2$ is then given by
\begin{equation}
\frac{d\Gamma}{dq^2}\,(\bar B^0\to \pi^+ \ell^- \bar\nu_\ell) = \frac{G_F^2
  |V_{ub}|^2}{ 192 \pi^3 m_B^3}\,\lambda^{3/2}(q^2) |f_+(q^2)|^2\,,
\end{equation}
where $\lambda(q^2) = (m_B^2+m_\pi^2-q^2)^2 - 4 m_B^2
m_\pi^2$ is the phase-space factor. The calculation of $f_+$ has
been the subject of numerous papers; the current state-of-the-art
methods are unquenched lattice simulations \cite{FNALno,HPQCD} and 
QCD sum rules on the light-cone (LCSRs) \cite{otherLCSR,BZ04}.
A particular challenge for any
theoretical calculation is the prediction of the {\em shape} of
$f_+(q^2)$ for all physical $q^2$: 
LCSRs work best for small $q^2$; lattice
calculations, on the other hand, are to date most reliable for large
$q^2$.
Hence, until very recently, the prediction of the
$B\to\pi \ell\nu$ decay rate necessarily involved an extrapolation of
the form factor, either to large or to small $q^2$. If, on the other
hand, the $q^2$ spectrum were known from experiment, the shape of $f_+$
could be constrained, allowing an extension of the LCSR and lattice
predictions beyond their region of validity. A first study of
the impact of the measurement, in 2005, of the $q^2$ spectrum in 5 bins in
$q^2$ by the BaBar collaboration \cite{bab} on the
shape of $f_+$ was presented in Ref.~\cite{BZvub}.
The situation has
improved dramatically in  2006 with the publication of
high-precision data of the $q^2$ spectrum
\cite{BaBarSL}, with 12 bins in $q^2$ and full statistical and
systematic error correlation matrices.
These data allow one to fit the
form factor to various parametrisations and determine the value of
$|V_{ub}| f_+(0)$ \cite{PB7}. As it turns out, the results from all but the
simplest parametrisation agree up to tiny differences which suggests
that the resulting value of $|V_{ub}| f_+(0)$ is {\em truly model-independent}.
In these proceedings we report the results for $|V_{ub}| f_+(0)$ and
$|V_{ub}|$, obtained in Ref.~\cite{BaBarSL,PB7}. 

There are four parametrisations of $f_+$ which are frequently used in
the literature. All of them include the essential feature that
$f_+$ has a pole at $q^2=m_{B^*}^2$; as $B^*(1^-)$ is a narrow
resonance with $m_{B^*}=5.325\,{\rm GeV}<m_B+m_\pi$, 
it is expected to have a
distinctive impact on the form factor. The parametrisations are:
\begin{itemize}
\item[(i)] Becirevic/Kaidalov (BK) \cite{BK}:
\begin{equation}\label{BK}
f_+(q^2) = \frac{f_+(0)}{\left(1-q^2/m_{B^*}^2\right)
  \left(1-\alpha_{\rm BK}\,q^2/m_{B}^2\right)}\,,
\end{equation}
where $\alpha_{\rm
  BK}$ determines the shape of $f_+$ and $f_+(0)$ the
  normalisation;
\item[(ii)] Ball/Zwicky (BZ) \cite{BZ04}:
\begin{eqnarray}
f_+(q^2) &=& f_+(0)\left(\frac{1}{1-q^2/m_{B^*}^2}\right.\nonumber\\
&&{}\left. + \frac{r
  q^2/m_{B^*}^2}{\left(1-q^2/m_{B^*}^2\right)\left(1-
\alpha_{\rm BZ}\,q^2/m_{B}^2\right)} \right),\label{BZ}
\end{eqnarray}
with the two shape parameters $\alpha_{\rm BZ}$, $r$ and the
normalisation $f_+(0)$; BK is a variant of BZ with $\alpha_{\rm BK} :=
\alpha_{\rm BZ} = r$;
\item[(iii)] the AFHNV parametrisation of Ref.~\cite{flynn}, based on
  an $(n+1)$-subtracted Omnes respresentation of $f_+$:
\begin{eqnarray}
f_+(q^2)& \stackrel{n\gg 1}{=}& \frac{1}{m_{B^*}^2-q^2}\,\prod_{i=0}^n \left[
  f_+(q_i)^2 (m_{B^*}^2 - q_i^2)\right]^{\alpha_i(q^2)}\\
&&\mbox{with}\quad \alpha_i(q^2) = \prod_{j=0,j\neq i}^n
  \frac{q^2-s_j}{s_i-s_j}\,;
\end{eqnarray}
the shape parameters are $f_+(q_i^2)/f_+(q_0^2)$ with
$q_{0,\dots n}^2$ the subtraction points; 
\item[(iv)] the BGL parametrisation based on the analyticity of $f_+$
  \cite{disper}: 
\begin{eqnarray}
f_+(q^2) & = & \frac{1}{P(t) \phi(q^2,q_0^2)}\,\sum_{k=0}^\infty
a_k(q_0^2) [z(q^2,q_0^2)]^k\,,\label{disper}\\
z(q^2,q_0^2) & = & \frac{\{m_+^2 - q^2\}^{1/2}
- \{m_+^2 - q_0^2\}^{1/2}}{ \{m_+^2 - q^2\}^{1/2}
+ \{m_+^2 - q_0^2\}^{1/2}}
\end{eqnarray}
with $m_+ = m_B + m_\pi$ and 
$\phi(q^2,q_0^2)$ as given in \cite{disper}. The ``Blaschke'' factor
$P(q^2) = z(q^2,m_{B^*}^2)$ accounts for the $B^*$ pole. The expansion
parameters $a_k$ are constrained by unitarity to fulfill $\sum_k a_k^2
\leq 1$. $q_0^2$ is a
free parameter that can be chosen to attain the tightest possible
bounds.
The series in (\ref{disper}) provides a systematic expansion in the
small parameter $z$, which for practical purposes has to be truncated
at order $k_{\rm max}$. 
The shape parameters are given by
$\{a_k\}$.
We minimize $\chi^2$ in $\{a_k\}$
for two choices of $q_0^2$: 
\begin{itemize}
\item[(a)] $q_0^2 = (m_B+m_\pi) (\sqrt{m_B}-\sqrt{m_\pi})^2 =
20.062\,{\rm GeV}^2$, which minimizes the possible values of $z$,
$|z|<0.28$, and hence also minimizes the truncation error of the series in
(\ref{disper}) across all $q^2$; the minimum $\chi^2$ is reached for 
$k_{\rm max} = 2$;
\item[(b)] $q_0^2 = 0\,{\rm GeV}^2$ with
$z(0,0)=0$ and $z(q^2_{\rm max},0)= -0.52$, which minimizes the
truncation error for small and moderate $q^2$ where the data are most
constraining; the minimum $\chi^2$ is reached for $k_{\rm max} = 3$. 
\end{itemize}
\end{itemize}
The advantage of BK and BZ is that they are both intuitive and simple;
BGL, on the other hand, offers a
systematic expansion whose accuracy can be adapted to that of the data
to be fitted, so we choose it as our default parametrisation.

We determine the best-fit parameters for all four parametrisations
from a minimum-$\chi^2$ analysis. 
\begin{table*}[t]
\renewcommand{\arraystretch}{1.3}
\addtolength{\arraycolsep}{3pt}
$$
\begin{array}{l||l|l}
& |V_{ub}| f_+(0) & \mbox{Remarks}\\\hline
{\rm BK} & (9.3\pm 0.3\pm 0.3)\times 10^{-4}
         & \chi^2_{\rm min}=8.74/11\,{\rm dof}\\
         & & \alpha_{\rm BK} = 0.53\pm 0.06\\\hline
{\rm BZ} & (9.1\pm 0.5\pm 0.3)\times 10^{-4}
         & \chi^2_{\rm min}=8.66/10\,{\rm dof}\\
         & & \alpha_{\rm BZ} 
= 0.40^{+0.15}_{-0.22},\,r = 0.64^{+0.14}_{-0.13} \\\hline
{\rm BGLa} & (9.1\pm 0.6 \pm 0.3)\times 10^{-4}
          & \chi^2_{\rm min}=8.64/10\,{\rm dof}\\
         && q_0^2 = 20.062\,{\rm GeV}^2\\
         && \theta_1 = 1.12^{+0.03}_{-0.04},\,\theta_2 = 4.45\pm 0.06\\\hline
{\rm BGLb} & (9.1\pm 0.6 \pm 0.3)\times 10^{-4}
          & \chi^2_{\rm min}=8.64/9\,{\rm dof}\\
         && q_0^2 = 0\,{\rm GeV}^2\\
         && \theta_1 = 1.41^{+0.02}_{-0.03},\,\theta_2 = 3.97\pm
  0.10\,,
         \theta_3 = 5.11^{+0.67}_{-0.39}\\\hline
{\rm AFHNV} & (9.1\pm0.3\pm0.3)\times 10^{-4} & \chi^2_{\rm min} = 8.64/8\,{\rm
  dof}\\
&& f_+(q^2_{\rm max}\cdot \{1/4,2/4,3/4,4/4\})/f_+(0)\\
&&  = \{1.54\pm
0.07,2.56\pm 0.11,5.4\pm 0.4,26\pm 11\}
\end{array}
$$
\caption[]{\small Model-independent results 
  for $|V_{ub}| f_+(0)$ using the BaBar data for the
  spectrum \cite{BaBarSL} and the HFAG average for the total branching
  ratio \cite{HFAG}. The first error comes from
  the uncertainties of the parameters determining the shape of $f_+$;
  these parameters are given in the right column; full definitions
can be found in Ref.~\cite{PB7}. The
  second error comes from the uncertainty of the branching ratio.
}\label{tab1}
\end{table*}
In Tab.~\ref{tab1} we give the results for $|V_{ub}|f_+(0)$ obtained from
fitting the various parametrisations to the BaBar data for the
normalised partial branching fractions in 12 bins of $q^2$:
$q^2\in \{[0,2],[2,4],[4,6],[6,8],[8,10],[10,12],[12,14],[14,16],$ $[16,18],
[18,20]$, $[20,22],[22,26.4]\}\,{\rm GeV}^2$; the absolute 
normalisation is gi\-ven by the HFAG average of the
semileptonic branching ratio, 
${\cal B}(\bar B^0\to \pi^+\ell^- \bar \nu_\ell) = (1.37\pm 0.06({\rm
  stat})\pm 0.07{\rm (syst)})\times 10^{-4}$ \cite{HFAG}. 
It is evident that good values of $\chi^2_{\rm min}$ are
obtained for all parametrisations. Our result is
\begin{equation}\label{16}
|V_{ub}|f_+(0) = (9.1\pm 0.6({\rm shape})\pm 0.3({\rm BR}))\times 10^{-4}
\end{equation}
from BGLa which we choose as default parametrisation.
We would like to stress that this result is {\em completely
  model-independent}, and also independent of the value of $|V_{ub}|$; it 
relies solely on the experimental data for $B\to\pi\ell\nu$ from BaBar
 for the spectrum and the HFAG average of the branching
  ratio. The BaBar collaboration finds \cite{BaBarSL} $|V_{ub}|f_+(0)
  = (9.6\pm 0.3({\rm stat})\pm 0.2({\rm syst}))\times 10^{-4}$, using
  the larger value of the branching ratio ${\cal B}(B\to\pi e\nu) =
  (1.46\pm 0.07\pm 0.08)\times 10^{-4}$, while \cite{flynn} quotes  
$|V_{ub}|f_+(0) = (8.8\pm 0.8)\times 10^{-4}$, based on a fit to all
  available data from BaBar, Belle, CLEO and form factor predictions
  from both light-cone sum rules and lattice calculations.

\begin{figure}[t]
\includegraphics[width=0.48\textwidth]{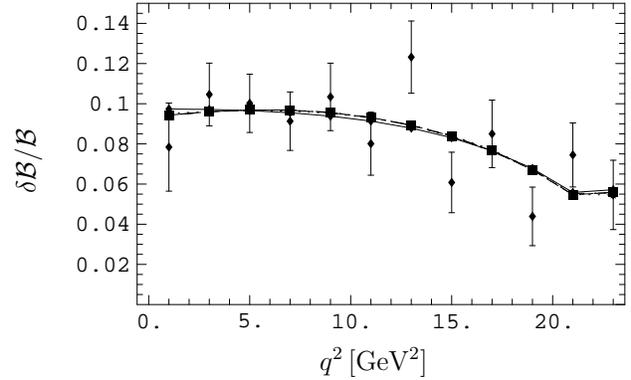}
\caption[]{\small Experimental data for the normalised branching ratio
  $\delta{\cal B}/{\cal B}$ per
  $q^2$ bin, $\sum \delta{\cal B}/{\cal B}=1$, and best
  fits.  The  lines are the best-fit results for
  the five different parametrisations listed in Tab.~\ref{tab1}.
The increase in the last bin is due to
  the fact that it is wider than the others ($4.4\,{\rm GeV}^2$ vs.\
  $2\,{\rm GeV}^2$).}\label{fig1}
\end{figure}

In  Fig.~\ref{fig1} we show the best-fit curves for all parametrisations
  together with the experimental data and error bars.
\begin{figure*}
\includegraphics[height=0.3\textwidth]{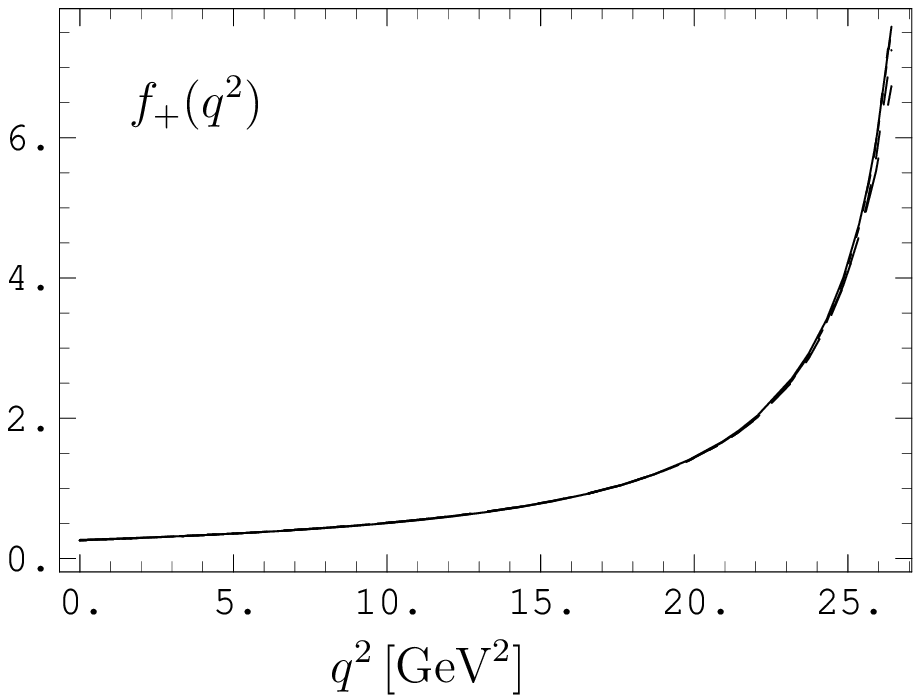}\hspace*{10pt}%
\includegraphics[height=0.3\textwidth]{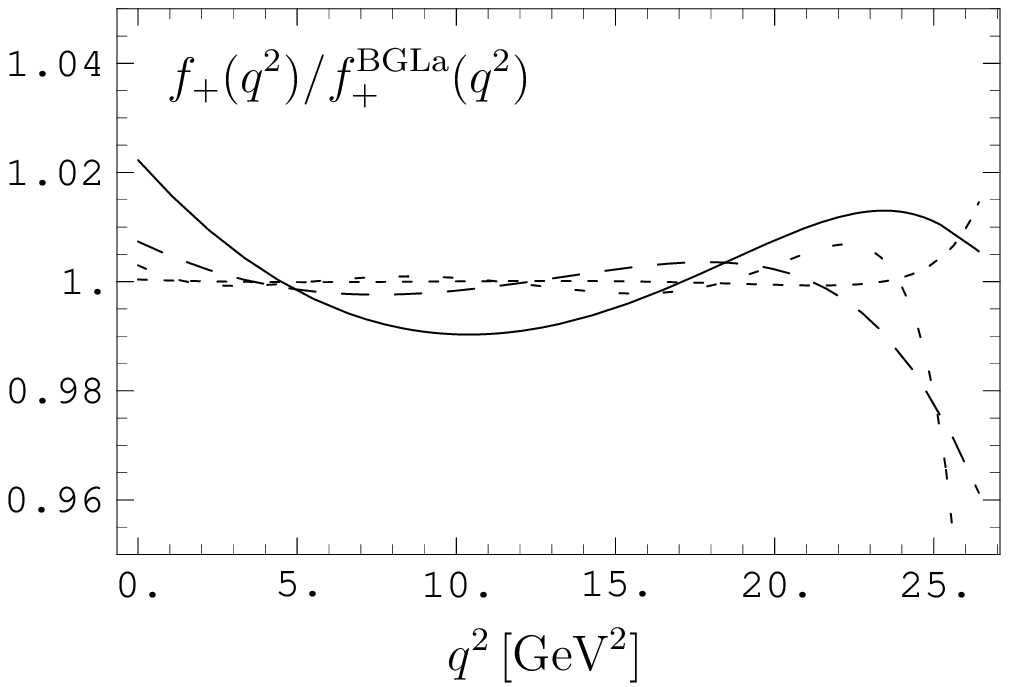}
\caption[]{\small Left panel: best-fit form factors $f_+$ as a function of
  $q^2$. The line is an overlay of all five parametrisations. 
Right panel: best-fit form factors normalised to BGLa. Solid
  line: BK, long dashes: BZ, short dashes: BGLb, short dashes with
  long spaces: AFHNV.}\label{fig2}
\end{figure*}
All fit curves basically coincide except for the BK parametrisation
which has a slightly worse $\chi^2_{\rm min}$. In Fig.~\ref{fig2} we show
the best-fit form factors themselves. The  curve in the left
panel is an overlay
of all five parametrisations; noticeable differences occur only for
large $q^2$, which is due to the fact that these points are 
phase-space suppressed in the spectrum and hence cannot be fitted with
high accuracy. In the right panel we graphically 
enhance the differences
between the best fits by normalising all parametrisations
to our preferred choice BGLa; for $q^2<25\,{\rm GeV}^2$, all best-fit form
factors agree within 2\%. 

\begin{table*}[tb]
\renewcommand{\arraystretch}{1.3}
\addtolength{\arraycolsep}{3pt}
$$
\begin{array}{l||l|l}
& \mbox{BK} & \mbox{BGLa}\\\hline
\mbox{LCSR}
& f_+(0)=0.26\pm 0.03\,,\quad \alpha_{\rm BK} = 0.63^{+0.18}_{-0.21}
& f_+(0)=0.26\pm 0.03\\
\mbox{Ref.~\cite{BZ04}}
& |V_{ub}| = (3.5\pm 0.6\pm 0.1)\times 10^{-4} & |V_{ub}| = (3.5\pm
0.4\pm 0.1)\times 10^{-4}\\
& |V_{ub}| f_+(0) = (9.0^{+0.7}_{-0.6}\pm 0.4)\times 10^{-4} &\\
\mbox{exp. input} & {\cal B}(B\to\pi \ell\nu)_{q^2\leq 16\,{\rm
    GeV}^2}
& {\cal B}(B\to\pi \ell\nu) \mbox{~and~BGLa}\\
&  = (0.95\pm 0.07)\times 10^{-4}
& \mbox{parameters from Tab.~\ref{tab1}}\\ \hline
\mbox{HPQCD} 
& f_+(0)=0.21\pm 0.03\,,\quad \alpha_{\rm BK} = 0.56^{+0.08}_{-0.11}
& f_+(0)=0.21\pm 0.03\\
\mbox{Ref.~\cite{HPQCD}} 
& |V_{ub}| = (4.3\pm 0.7\pm 0.3)\times 10^{-4} & |V_{ub}| = (4.3\pm
0.5\pm 0.1)\times 10^{-4}\\
& |V_{ub}| f_+(0) = (8.9^{+1.2}_{-0.9}\pm 0.4)\times 10^{-4} &\\
\mbox{exp. input} & {\cal B}(B\to\pi \ell\nu)_{q^2\geq 16\,{\rm
    GeV}^2}
& {\cal B}(B\to\pi \ell\nu) \mbox{~and~BGLa}\\
&  = (0.35\pm 0.04)\times 10^{-4} 
& \mbox{parameters from Tab.~\ref{tab1}}\\ \hline
\mbox{FNAL} 
& f_+(0)=0.23\pm 0.03\,,\quad \alpha_{\rm BK} = 0.63^{+0.07}_{-0.10}
& f_+(0)=0.25\pm 0.03\\
\mbox{Ref.~\cite{FNALno}} 
& |V_{ub}| = (3.6\pm 0.6\pm 0.2)\times 10^{-4} & 
|V_{ub}| = (3.7\pm 0.4\pm 0.1)\times 10^{-4}\\
& |V_{ub}| f_+(0) = (8.2^{+1.0}_{-0.8}\pm 0.3)\times 10^{-4} &\\
\mbox{exp. input} & {\cal B}(B\to\pi \ell\nu)_{q^2\geq 16\,{\rm
    GeV}^2}
& {\cal B}(B\to\pi \ell\nu)\mbox{~and~BGLa}\\
&   = (0.35\pm 0.04)\times 10^{-4} 
& \mbox{parameters from Tab.~\ref{tab1}}
\end{array}
$$
\caption[]{\small $|V_{ub}|$ and $|V_{ub}| f_+(0)$ from various theoretical
  methods. The column labelled BK gives the results obtained from a
  fit of the form factor to the BK parametrisation, and the column
  labelled BGLa those from a fit of $f_+(0)$ 
to the best-fit BGLa parametrisation from Tab.~\ref{tab1}. The first
  uncertainty comes from the shape parameters, the second from the
  experimental branching ratios;
  the latter are taken from HFAG \cite{HFAG}.
}\label{tab4}
\end{table*}

As mentioned above, theoretical predictions for $f_+$ are available
from lattice calculations and LCSRs. The LCSR calculation \cite{BZ04} 
includes twist-2 and -3 contributions to $O(\alpha_s)$ accuracy and twist-4
contributions at tree-level. The lattice calculations
\cite{FNALno,HPQCD} are unquenched 
with $N_f=2+1$ dynamical flavours, i.e.\
mass-degenerate $u$ and $d$ quarks and a heavier $s$ quark.
The obvious questions are (a) whether these
predictions of $f_+(q^2)$
are compatible with the experimentally determined shape of
the form factor and (b) what the resulting value of $|V_{ub}|$
is. In order to answer these questions, we follow two different
procedures. We first fit the lattice and LCSR
form factors to the BK parametrisation and extract $|V_{ub}|$, for
lattice, from ${\cal B}(B\to\pi\ell\nu)_ {q^2\geq 16\,{\rm
  GeV}^2}$, and, for LCSRs, from ${\cal B}(B\to\pi\ell\nu)_ {q^2\leq 16\,{\rm
  GeV}^2}$; the cuts in $q^2$ are imposed in order to minimise any
uncertainty from extrapolating in $q^2$. 
The results are shown in the BK column of
Tab.~\ref{tab4}. Equipped with the experimental information on the form factor
shape, i.e.\ the BGLa parametrisation of Tab.~\ref{tab1},
we also follow a different procedure and 
perform a fit of the theoretical predictions to
this shape, with the normalisation $f_+(0)$ as fit parameter. The corresponding
results are shown in the right column. 
Comparing the errors for $|V_{ub}|$ in both columns,
  it is evident that the main impact of the experimentally fixed
  shape, i.e.\ using the BGLa parametrisation of $f_+$,
  is a reduction of both theory and experimental errors; this is due to the
  fact that, once the shape is fixed, $|V_{ub}|$ can be determined
  from the full branching ratio with only 3\% experimental
  uncertainty, whereas the partical branching fractions in the BK
  column induce 4\% and 6\% uncertainty, respectively, for $|V_{ub}|$;
  the theory error gets reduced because the theoretical uncertainties
  of $f_+$ predicted for various $q^2$ are still rather large, which 
  implies theory uncertainties
  on the shape parameter $\alpha_{BK}$, which are larger than those of
  the experimentally fixed shape parameters.

What is the conclusion to be drawn from these results? Let us compare
with $|V_{ub}|$ from inclusive determinations.
HFAG gives results obtained using dressed-gluon exponentiation (DGE)
\cite{einan} and the shape-function formalism (BLNP) \cite{neubert}:
\begin{eqnarray}
|V_{ub}|_{\rm incl,DGE}^{\rm HFAG} &=& (4.46\pm 0.20\pm
 0.20)\times 10^{-3}\,,\nonumber\\
|V_{ub}|_{\rm incl,BLNP}^{\rm HFAG} &=& (4.49\pm 0.19\pm
 0.27)\times 10^{-3}\,,\label{VubHFAG}
\end{eqnarray}
where the first error is experimental (statistical and systematic) and
the second external (theoretical and parameter uncertainties). Both
results are in perfect agreement. Note that, at this conference,
Neubert has quoted a lower value, $|V_{ub}| = (4.10\pm 0.30({\rm exp})
\pm 0.29({\rm th}))\times 10^{-3}$ \cite{n2}, based on the subclass of
``best'' determinations (with highest efficiency and best theoretical control).
At the same time, $|V_{ub}|$ can also be determined in a more
indirect way, based on global fits of the unitarity triangle (UT),
using only input from various CP violating observables which are
sensitive to the angles of the UT. 
Following the UTfit collaboration, we call the corresponding fit of UT
parameters UTangles. Both the UTfit \cite{UTfit} and the
CKMfitter collaboration \cite{CKMfitter,private} find
\begin{equation}\label{VubUT}
|V_{ub}|_{\rm UTangles}^{\rm UTfit,CKMfitter} = (3.50\pm 0.18)\times
 10^{-3}\,.
\end{equation}
The discrepancy between (\ref{VubHFAG}) and (\ref{VubUT}) starts to
become significant. One interpretation of this result is that there is
new physics (NP) in $B_d$ mixing which impacts the value of $\sin 2\beta$
from $b\to ccs$ transitions, the angle measurement with the
smallest uncertainty. The value of $|V_{ub}|$ in (\ref{VubHFAG}) implies
\begin{equation}
\left.\beta\right|_{|V_{ub}|_{\rm incl}^{\rm HFAG}} = (26.9\pm
2.0)^\circ\, \leftrightarrow \,\sin 2\beta = 0.81\pm 0.04\,,
\end{equation}
using the recent Belle result $\gamma=(53\pm 20)^\circ$ from the
Dalitz-plot analysis of the tree-level process $B^+\to D^{(*)}K^{(*)+}$ 
\cite{Bellegamma}.\footnote{See Refs.~\cite{UTfit,CKMfitter,vtdts,yetagain} 
for alternative
  determinations of $\gamma$.}
  This value disagrees by more than $2\sigma$ 
with the HFAG average for $\beta$ from 
$b\to ccs$ transitions, $\beta = (21.2\pm 1.0)^\circ$ ($\sin 2\beta =
0.675\pm 0.026$). The difference
  of these two results indicates the possible presence of a NP phase in
  $B_d$ mixing, $\phi_d^{\rm NP} \approx -10^\circ$.
This interpretation of the experimental situation 
is in line with that of Ref.~\cite{newphase}.
An alternative 
interpretation is that there is actually no or no significant NP in the
mixing phase of $B_d$ mixing, but that the uncertainties in either
UTangles or inclusive $b\to u\ell\nu$ transitions (experimental and
theoretical) or both are underestimated and that (\ref{VubHFAG}) and 
(\ref{VubUT}) actually do agree.
The main conclusion from this discussion is that both LCSR and FNAL
predictions for $f_+$ support the UTangles value for $|V_{ub}|$,
and differ at the $2\sigma$ level from the inclusive $|V_{ub}|$,
whereas HPQCD supports the inclusive result. Using the experimentally
fixed shape of $f_+$ in the analysis instead of fitting it to the
theoretical input points reduces both the theoretical and experimental
uncertainty of the extracted $|V_{ub}|$.

To summarize, we have presented a truly model-independent 
determination of the quantity $|V_{ub}| f_+(0)$ from the experimental data
for the spectrum of $B\to\pi\ell\nu$ in the invariant lepton mass
provided by the BaBar collaboration \cite{BaBarSL}; 
our result is given in (\ref{16}). We
have found that the BZ, BGL and AFHNV parametrisations of the form
factor yield, to within 2\% accuracy, the same results for
$q^2<25\,{\rm GeV}^2$. We then have used the best-fit BGLa shape of $f_+$ to
determine $|V_{ub}|$ using three different theoretical predictions for
$f_+$, QCD sum rules on the light-cone \cite{BZ04}, and the lattice
results of the HPQCD \cite{HPQCD} and FNAL collaborations
\cite{FNALno}. The advantage of this procedure compared to that
employed in previous works, where the shape was determined from the
theoretical calculation itself, is a
reduction of both experimental and theoretical uncertainties of the
resulting value of $|V_{ub}|$. We have found that the LCSR and FNAL
form factors yield values for $|V_{ub}|$ which
agree with the UTangles result, but
differ, at the $2\sigma$ level, from the HFAG value
obtained from inclusive decays. The HPQCD form factor, on
the other hand, is compatible with both UTangles and the inclusive
$|V_{ub}|$. Our results show a certain preference for the UTangles
result for $|V_{ub}|$, disfavouring a new-physics scenario in $B_d$
mixing, and highlight the need for a
re-analysis of $|V_{ub}|$ from inclusive $b\to u \ell\nu$ deacys.

\subsection*{Acknowledgements}
This work was supported in part by the EU networks
contract Nos.\ MRTN-CT-2006-035482, {\sc Flavianet}, and
MRTN-CT-2006-035505, {\sc Heptools}.

\end{document}